# A Multivocal Literature Review on the Benefits and Limitations of Automated Machine Learning Tools

Kelly Azevedo[a], Luigi Quaranta[b], Fabio Calefato[b], Marcos Kalinowski[a]

[a]*PUC-Rio, Rio de Janeiro, Brazil*
[b]*University of Bari, Bari, Italy*

**Abstract**

***Context***. Rapid advancements in Artificial Intelligence (AI) and Machine Learning (ML) are revolutionizing software engineering in every application domain, driving unprecedented transformations and fostering innovation. However, despite these advances, several organizations are experiencing friction in the adoption of ML-based technologies, mainly due to the current shortage of ML professionals. In this context, Automated Machine Learning (AutoML) techniques have been presented as a promising solution to democratize ML adoption, even in the absence of specialized people.

***Objective***. Our research aims to provide an overview of the evidence on the benefits and limitations of using AutoML tools.

***Method***. We conducted a Multivocal Literature Review (MLR), which allowed us to identify 54 sources from the academic literature and 108 sources from the grey literature reporting on AutoML benefits and limitations. We extracted explicitly reported benefits and limitations from the papers and applied the thematic analysis method for synthesis.

***Results***. Overall, we identified 18 reported benefits and 25 limitations. Concerning the benefits, we highlight that AutoML tools can help streamline the core steps of ML workflows, namely data preparation, feature engineering, model construction, and hyperparameter tuning—with concrete benefits on model performance, efficiency, and scalability. In addition, AutoML empowers both novice and experienced data scientists, promoting ML accessibility. On the other hand, we highlight several limitations that may represent obstacles to the widespread adoption of AutoML. For instance, AutoML tools may introduce barriers to transparency and interoperability, exhibit limited flexibility for complex scenarios, and offer inconsistent coverage of the ML workflow.

***Conclusions***. The effectiveness of AutoML in facilitating the adoption of machine learning by users may vary depending on the specific tool and the context in which it is used. As of today, AutoML tools are used to increase human expertise rather than replace it, and, as such, they require skilled users.

## 1. Introduction

The domain of technology has experienced significant advances in Artificial Intelligence (AI) and Machine Learning (ML), which have invaded diverse industries such as automotive, business, and healthcare [1, 2, 3]. The widespread adoption of machine learning methodologies has significantly transformed data analysis and decision-making processes in various industries.

Despite these advances, numerous organizations struggle to implement vital machine learning initiatives, frequently relying heavily on specialized expertise [4, 5, 6]. This dependency creates an obstacle, preventing companies from fully utilizing machine learning and achieving their desired results, as the demand for machine learning professionals has increased [7]. The current situation becomes worse because there are not enough skilled people trained in advanced machine learning techniques. This further limits industrial progress and restricts innovation.

AutoML provides a potential approach to reduce the dependence on machine learning professionals and enable organizations to use machine learning efficiently [8]. AutoML tools typically streamline several elements of ML project development, including data cleansing, feature and model selection, hyperparameter tuning, and other features. AutoML aims to optimize these procedures, improving efficiency and offering faster outcomes than manually devised approaches. In this sense, AutoML software solutions can be crucial for driving machine learning progress by facilitating organizations in optimizing the use of ML capabilities.

Leading IT giants such as AWS[1], Google[2], IBM[3], and Microsoft[4] have acknowledged the growing business need for AutoML technology and have created their own AutoML platforms. The increased focus highlights the importance of AutoML as a disruptive technology that can change the ML development and deployment process.

In this work, we aim to explore the evidence on AutoML technologies and reveal the reported benefits and limitations. To this end, we conducted a Multivocal Literature Review (MLR), i.e., a form of systematic literature review that includes the analysis of non-peer-reviewed articles along with the academic literature. In particular, this study aims to contribute by identifying:

- The benefits revealed in scientific investigations on AutoML and the perceived benefits being reported by practitioners and companies that have embraced AutoML tools according to grey literature sources.

- The limitations revealed in scientific investigations and the perceived limitations being reported by practitioners and companies when using AutoML tools according to grey literature sources.

To the best of our knowledge, this is the first secondary study on AutoML to apply the MLR methodology. AutoML, as an emerging technology, has only recently garnered attention from researchers. By complementing scientific evidence with additional insights from non-peer-reviewed articles, such as technical blogs, we have been able to present a nuanced and thorough analysis of its reported benefits and limitations, encompassing a wide range of perspectives.

The remainder of this paper is structured as follows. In Section 2, we briefly summarize the existing landscape of ML and AutoML. In Section 3, we describe the protocol followed to carry out our multivocal literature review of articles on AutoML. In Section 4 we present the study results, which are then discussed in Section 5. Finally, in Sections 6 and 7, we present the limitations of our research and draw conclusions.

---

[1]https://aws.amazon.com/machine-learning/automl/
[2]https://cloud.google.com/automl
[3]https://www.ibm.com/topics/automl
[4]https://learn.microsoft.com/azure/machine-learning

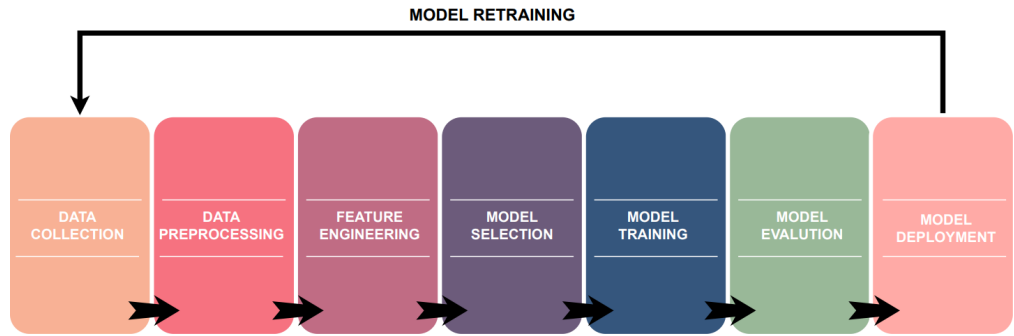

Figure 1: Simplified Machine Learning workflow representation (adapted from Amershi *et al.* ML 9 steps workflow [9]).

## 2. Background and related work

Industries worldwide are seeing a significant shift, driven by the rapidly growing science of machine learning [1, 2, 3]. This technology has enabled the possibility of automation, optimization, and data-driven insights, enabling computers to acquire knowledge and make predictions or judgments without explicit programming. This section will delve into the concepts of machine learning workflow and automated machine learning and summarize the pertinent existing research in this field that is related to this study.

### 2.1. Machine Learning Workflow

Machine Learning (ML) workflows are systematic processes that transform raw data into trained models capable of making informed predictions or decisions. These workflows are key to numerous applications, and a typical ML workflow consists of several key steps. Amershi *et al.* [9] presents the nine steps of machine learning workflow, including both data-centric (collecting, cleaning, labeling) and model-centric (requirements, feature engineering, training, evaluation, deployment, monitoring) steps. The study highlights the importance of many feedback loops, suggesting that the assessment and surveillance of models might impact any previous step.

Based on the nine ML life cycle stages presented by Amershi *et al.* [9] and the CRISP-DM industry-independent process model phases [10], we abstracted seven generic life cycle stages (Figure 1), similar to what was done by [11]. The background explanations hereafter are based on the textbook explanations contained in [11].

Data collection and data preprocessing are the initial steps. Data is collected, cleaned, and prepared for analysis (*e.g.*, labeled, transformed) during these steps. The quality of data in this phase can significantly impact the final model's performance. Feature engineering is a creative and knowledge-intensive step that follows data preprocessing. It involves selecting, transforming, and even creating features from the dataset. This process enhances the model's ability to make accurate predictions by making the data more informative and relevant.

The next critical step is model selection, where the ML algorithm that is best suited to the problem is chosen. The decision takes into account the nature of the data and the specific task, whether it is classification, regression, clustering, or another application. The chosen algorithm establishes the foundation for the entire workflow.

With the algorithm selected, the model training step begins. Here, the algorithm learns patterns and relationships within the data; it adjusts its internal parameters to better represent the underlying patterns in the data. The next step is model evaluation, which is essential to understanding how well the model is likely to perform on unseen data. Various metrics, depending on the problem (*e.g.*, accuracy, precision, recall, or F1 score could be used for classification problems while the mean squared error and R-squared could be used for regression problems), are used to assess the model's performance.

Depending on the model's performance, it can be improved by adjusting the hyperparameters. The model does not learn hyperparameters, and they must be configured before training. This step, called hyperparameter tuning, is an iterative process to optimize the hyperparameters of the model. This fine-tuning step optimizes the model's performance, and techniques such as grid search or random search are commonly employed.

Finally, after successful training and validation, the model is ready for deployment in a real-world setting. Model deployment can take the form of APIs, embedded systems, or cloud-based services, enabling it to make predictions or automate decisions.

ML workflows come with a set of challenges. Nahar *et al.* [12] collected and organized overall challenges related to building products with ML-components, some of which are directly related to the ML workflow itself. For instance, data quality is a critical issue, as poor quality data can significantly impact model performance, introducing biases, inaccuracies, and decreasing overall accuracy. Furthermore, effective feature engineering requires domain expertise and an in-depth understanding of the data, making it a complex and creative process. Model selection can also be challenging, as different algorithms perform differently based on the data and task, requiring a deep understanding of the field. Balancing model complexity and generalization is a common challenge, as complex models may overfit training data, while simple ones may underfit, both resulting in poor generalization. Additionally, privacy and security concerns may arise, particularly when handling sensitive information. Scalability may become an issue as datasets grow, requiring additional computational resources. Moreover, the interpretability of models may be crucial, especially in sensitive applications like healthcare or finance, as complex "black-box" models can be difficult to explain to stakeholders or regulatory bodies.

In summary, ML workflows encompass well-defined phases, each associated with its particular set of challenges. Effectively addressing these challenges is essential to successfully deploy ML solutions in real-world applications and derive meaningful insights from the data.

*2.2. Automated Machine Learning*

Automated machine learning (AutoML) solutions have emerged as a response to the growing demand for machine learning, aiming at simplifying and expediting the model generation process.

AutoML solutions have significantly streamlined the complex landscape of machine learning processes. These tools promise to empower users to build high-quality machine learning models with minimal manual effort, overseeing everything from data pre-processing to hyperparameter optimization [13, 14, 15, 16, 17, 18]. AutoML technologies aim to democratize the field by making machine learning more accessible, even to individuals with limited coding or data science experience [19, 16, 20, 15, 21, 22, 23, 24].

They execute these tasks using a blend of statistical methods and optimization algorithms, which not only save time, but also diminish the necessity for manual intervention. Additionally, some AutoML tools extend their offerings to include features such as model deployment, scaling efficiency [25, 15], and improving efficiency and productivity [19, 13, 16, 14, 23].

The wide variety of AutoML tools and the range of user needs they address are clear indicators of their adaptability. Some tools focus on providing APIs and libraries that work well with R and Python, while others provide graphical user interfaces to enable creating machine learning models without scripting. Open-source AutoML frameworks also give users a lot of freedom to adapt to their needs. Commercial AutoML solutions introduce advanced features, including enhanced scalability and automated model deployment.

By automating routine tasks and allowing the potential elevation of model performance, AutoML tools have the potential to change the game when it comes to machine learning. They are constantly improving and adding new features to allow organizations and individuals to take advantage of machine learning capabilities.

*2.3. Related work*

Our multivocal literature review is a significant effort to provide a comprehensive overview of the field of AutoML research, setting itself apart from previous studies by its systematic protocol. After a careful examination of various important articles in the field of AutoML, it is clear that our study contributes by consolidating key advantages and difficulties of AutoML. In the next paragraphs, we will look at the related work in the area of AutoML.

Elshawi *et al.* [26] provide a useful background on difficulties with AutoML execution. The article highlights key practical challenges encountered when deploying AutoML solutions, emphasizing scalability, optimization techniques, time budgeting, and user-friendliness. Although they mainly aim to analyze current problems comprehensively, their work does not adhere to the structure of a systematic literature review.

Escalante [27] presents the history of AutoML and classifies its development into different waves. Although clear, this historical perspective mostly shows how methodologies have changed over time. They also did not perform a systematic review approach to compile the results of various AutoML studies.

Prado and Digiampietri [28] present a systematic review of automated feature engineering solutions in machine learning problems. With the main objective of identifying and analyzing the existing methods and techniques to perform the automated feature engineering step within the framework of machine learning problems, this study is written in Portuguese and focuses on feature engineering and not on the whole ML pipeline.

Thirunavukarasu *et al.* [29] compile a list of the clinical uses of AutoML, look at the strengths and weaknesses of the platforms that were used, judge the reliability of the research that tested AutoML, and compare the performance of these platforms with models traditionally created. However, their focus is limited to healthcare.

Branco *et al.* [30] examine current discussions on AutoML methods for electrical biosignal problems by analyzing articles published in six databases focused on technology and machine learning between 2018 and 2022. As a result, they outline the current difficulties in the field, provide insights on biosignal understanding, and identify the best AutoML solutions. However, their focus also is limited to healthcare.

Baymurzina *et al.* [31] examine the most current research on Neural Architecture Search (NAS), which represents a very specialized field of AutoML tools, and highlight various important concepts and issues that are associated with this topic. Hence, their focus is specifically set on NAS algorithms and tools.

Marques *et al.* [32] aim to discover and evaluate AutoML research in the context of multi-label classification and multi-target regression through a systematic literature review (SLR). However, their research questions do not involve the benefits and limitations of AutoML tools as a whole.

Nagarajah and Poravi [33] review the current state of AutoML, hyperparameter tuning, and meta-learning. They analyze many methods and evaluate them based on the algorithms they support, the features they provide, and how well they work in practice. However, they did not use a systematic approach when doing the research.

Wen and Li [34] answer questions related to the benefits and limitations of AutoML, but in the area of spatial decision support systems. The main goal of their paper is to analyze the benefits of using AutoML tools in spatial decision support systems.

Khalid *et al.* [35] conduct a systematic literature review of the challenges related to declarative machine learning AutoML solution (which allow users to express their intents through high-level abstractions). They included only 'white' literature (*i.e.*, official peer-reviewed articles published in academic journals or conference proceedings) published until May 2022. They used a database-only search strategy, excluded AutoML solutions that were not declarative and provided limited details on their analysis procedures.

Barbudo *et al.* [36] review the literature on AutoML from 2014 to 2021. The exclusion criteria filtered out papers lacking clear evidence of a blind, peer-review process and those not published in conferences ranked A* or A by the CORE ranking system, as well as papers from non-JCR indexed journals. This study has four research questions. Initially, it seeks to identify commonly used AutoML terms from original research. Secondly, the article takes a quantitative look at the research trajectory inside AutoML to see how it has evolved. Third, it explores the various phases of the knowledge discovery process covered by different AutoML tasks and the various techniques employed. Finally, the study identifies emerging trends and unexplored areas, thereby pinpointing potential paths for future research within AutoML. Their study does not focus on revealing reported benefits and limitations.

The literature review of the paper at hand represents an important step forward in the field of AutoML. Using a systematic approach, it thoroughly examines the advantages and disadvantages of AutoML, from sources between 2017 and 2022, synthesizing a wide range of academic and grey literature in the process. Beyond focusing on the specifics, our review provides a comprehensive analysis of the AutoML field as a whole. The analysis of the challenges and advancements in AutoML is made possible by consolidating findings from many sources. This multivocal literature review addresses a significant knowledge gap in the literature to help researchers, practitioners, and industries understand the main benefits and limitations of AutoML.

## 3. Methodology

### 3.1. Goal and Research Questions

This study aims to analyze the usage of AutoML tools and catalog their key advantages, challenges, and limitations. Accordingly, the research questions addressed by this study are:

**RQ1:** *What are the main benefits of AutoML tools?*

**RQ2:** *What are the main limitations of AutoML tools?*

To address RQ1 and RQ2, we conducted a multivocal literature review [37]. The result of this literature analysis is a comprehensive catalog that highlights the most frequently cited drawbacks and advantages of using AutoML tools.

### 3.2. Multivocal Literature Review

A multivocal literature review (MLR) is a systematic review that includes both official peer-reviewed articles published in academic journals or conference proceedings (referred to as 'white' literature) and informal documents from the 'grey' literature, such as whitepapers or blog posts, which are created by professionals, mainly based on their practical expertise.

Garousi *et al.* [37] demonstrated advantages of augmenting Systematic Literature Reviews (SLRs) with grey literature in software engineering research. Software engineers mostly depend on grey literature to stay informed. Additionally, they often disseminate their thoughts and experiences in the form of grey literature for similar purposes. Researchers fail to capture a significant amount of valuable information when they restrict their investigations to analyzing only the literature produced by academic authors.

Assuming that a similar situation must also take place in the domain of machine learning, we decided to include the analysis of grey literature in our research. In the following sections, we comprehensively explain the procedures involved in carrying out our Multivocal Literature Review (MLR).

### 3.3. MLR protocol

In May 2023, we searched for articles in three distinct search engines. To obtain scientific publications, we precisely applied the strategy described by Wohlin *et al.* [38], which has shown successful in identifying relevant primary studies in several investigations [39, 38]. This strategy consists of conducting a database search on Scopus[1], applying the study selection process to retrieved candidate papers to gather a fair and representative seed set, and then complementing the primary study identification with iterative snowballing until saturation is reached. To get the grey literature, we conducted an online search using the Google[2] search engine and the Gartner[3] knowledge database.

Hereafter, we provide detailed description of the entire MLR protocol, including the search strategy, selection criteria, data collection, data extraction, and data analysis.

#### 3.3.1. Search strategy

Our multivocal literature review adopted two distinct search approaches to cover both white and grey literature. These approaches were designed to ensure a thorough exploration of the investigated topic. In the following, we outline the specific strategies we used for each type of literature.

---

[1]https://www.scopus.com/
[2]https://www.google.com/
[3]https://www.gartner.com

**Search strategy for the white literature**

For the white literature, we followed the search strategy combining database searches with snowballing described by Wohlin *et al.* [38]. We defined a search string to be applied on Scopus in order to retrieve a representative seed set for snowballing.

To structure our search, we used the PICO framework [40], which categorizes search terms into *Population*, *Intervention*, *Comparison*, and *Outcome*. For our study, the population of interest concerns investigations related to AutoML, and the intervention concerns the benefits and limitations. We did not consider specific comparisons or outcomes but rather intended to analyze the benefits and limitations qualitatively. Our final search query for the white literature is represented in Figure 2.

> ("automl")
> **AND**
> ("benefits" **OR** "challenges" **OR** "limitations")

Figure 2: Final search query for the white literature.

This query was intentionally kept concise, as we planned to make use of forward and backward snowballing iterations in later stages [38]. For the first step of the search strategy (to create the snowballing seed set), we used the Scopus database.

**Search strategy for the grey literature**

For the grey literature, we used Google Search and crafted a more comprehensive search string. Since it is complex and often unfeasible to apply the snowballing technique to the grey literature, we aimed to broaden our search by including other synonyms and considering the AutoML leaders identified in the Gartner Magic Quadrant for Cloud AI Developer Services [41], namely Microsoft, Google, IBM, and AWS.

The final search query for the grey literature can be seen in Figure 3. This comprehensive search string ensured that we considered a wide range of terms, including synonyms, and limited our search to articles mentioning at least one of the key AutoML vendors to meet our research criterion.

```
("automl" OR "automated machine learning" OR "automatic machine learning")
                                    AND
        ("benefits" OR "advantages" OR "pros" OR "strengths" OR
    "opportunities" OR "challenges" OR "limitations" OR "issues" OR
       "difficulties" OR "cons" OR "disadvantages" OR "pitfalls" OR
              "downsides" OR "weaknesses" OR "drawback")
                                    AND
                   (Microsoft OR Google OR IBM OR AWS)
```

Figure 3: Final search query for the grey literature.

To improve our grey literature, we also conducted a search on Gartner database, a well-known research and advisory firm. In this database, we used a very simple query string, "automl", because we noticed that the number of results did not change when we used a more detailed query search.

By implementing these distinct search strategies, we aimed to create a well-rounded and exhaustive review of both white and grey literature on the benefits and limitations of AutoML tools.

*3.3.2. Selection criteria*

The exclusion (EC) criteria are crucial in refining the selection of documents for a Multivocal Literature Review. The ones we adopted are reported in Table 1, along with their rationale.

There are many different tools in the field of Automated Machine Learning (AutoML), ranging from free to high-cost ones. These tools include several different programming languages and frameworks, each with its own features that are useful at different points in the machine-learning workflow.

We decided to concentrate our studies on the leaders of this field, so we took advantage of the Gartner Magic Quadrant, which is both a research method and a graphic representation to rank companies in certain technology markets. It rates vendors based on how well they understand and can implement that vision in a certain industry or sector.

In the Magic Quadrant, vendors are put in a two-dimensional grid with their ability to execute on the y-axis and their completeness of vision on the x-axis. There are four groups: Leaders, Challengers, Visionaries, and Niche Players. Each quadrant shows a different part of a seller's position in the market. The leaders are vendors who usually have a clearly defined vision

**Search Results Exclusion Criteria**

| | Criterion | Rationale |
|---|---|---|
| **EC1** | Not published between 2017 and 2022. | This criterion narrows the literature down to recent sources, ensuring relevance and up-to-date information. |
| **EC2** | Not written in English. | This ensures that the literature reviewed is in a language accessible to most researchers. |
| **EC3** | Not a blog post, article, conference paper, or book chapter. | This restriction ensures that the documents we include in our review are of sufficient academic or professional quality. |
| **EC4** | Document not available for our institutions. | This criterion helps in streamlining the literature review process by focusing on sources that the researchers can readily access. |
| **EC5** | It does not include Google, Amazon, Microsoft, or IBM AutoML tools. | This specific criterion ensures that the literature reviewed is aligned with the major AutoML vendors, as per the Gartner Magic Quadrant for Cloud AI Developer Services. |
| **EC6** | Other AutoML-related publications that do not emphasize AutoML benefits, challenges, and limitations. | Our focus is on the benefits, challenges, and limitations of using AutoML tools. |

Table 1: The exclusion criteria and their rationale.

and are good at putting it into action. People see them as market leaders because they consistently set new standards and develop new ideas. In our research, we decided to narrow our investigation into the sources that would refer to the four leaders of Cloud AI Developer Services [41]: Google, AWS, IBM, and Microsoft, as can be seen in Figure 4.

*3.3.3. Data collection*

For the white literature, we applied the query search in the Scopus database and applied the exclusion criteria. From the remaining articles, we started applying forward and backward snowballing. We applied a hybrid approach (i.e., a combination of database search and snowballing technique) involving backward and forward snowballing iterations to curate our selection of relevant studies. As mentioned by [38], the hybrid search strategy is of significant importance in systematic literature studies due to its effectiveness.

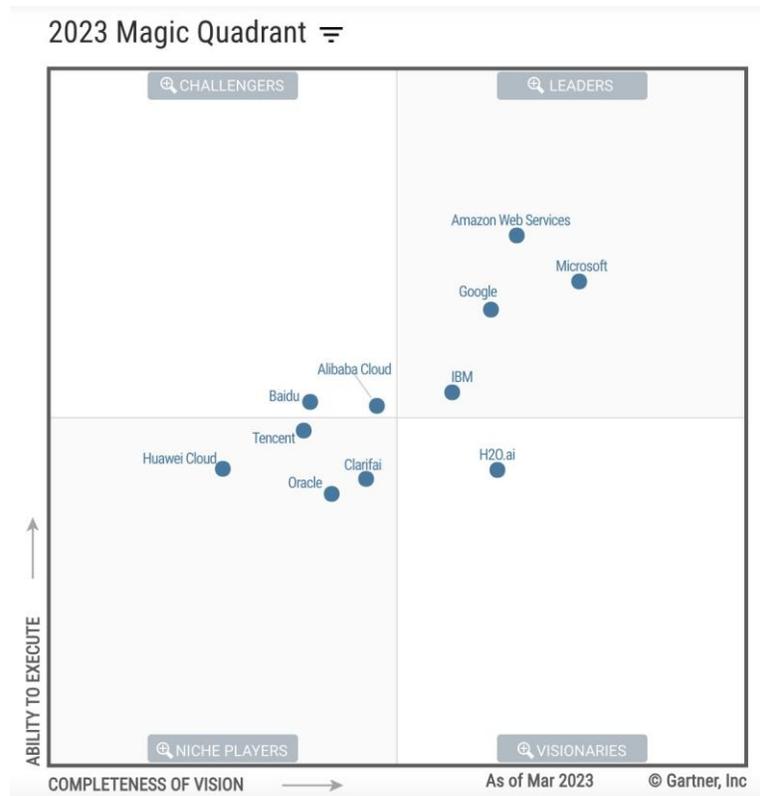

Figure 4: Gartner Magic Quadrant for Cloud AI Developer Services [41].

Our approach extended beyond Scopus, encompassing Semantic Scholar[1], and Google Scholar[2], for the snowballing process. We started by using Google Scholar for citation retrieval, as it consistently offered a more extensive range of citations than Scopus and Semantic Scholar. After the first round of snowballing, we transitioned to Semantic Scholar because it allows data access through an API, which streamlined the data collection process.

In total, we conducted three rounds of snowballing. We concluded our search at the third round as no additional articles that met our criteria could be identified, reaching snowballing saturation. Our initial pool of articles comprised approximately 5,700 papers, which can be found in our online open science repository [42]. Before implementing our exclusion criteria, we performed a validation check on the papers. This entailed checking for duplicates, inspecting references classified as grey literature, and verifying that references or citations contained the keywords "automl", "automated machine learning" or "automatic machine learning" within their titles, abstracts, or keywords.

---

[1] https://www.semanticscholar.org/
[2] https://scholar.google.com/

During the snowballing process, we excluded grey literature, recognizing that a separate Google Search was concurrently conducted to address this specific category of sources. Additionally, to maintain the focus on AutoML, we eliminated papers that did not feature the designated keywords in their title, abstract, or keywords. This decision was motivated by the observation that several articles under analysis pertained to general Machine Learning topics rather than Automated Machine Learning.

For the grey literature, we employed the Google Search query outlined in Section 3.3.1. Upon analyzing the results, we evaluated the exclusion criteria and determined that we would stop the search after three consecutive pages – each containing ten results – with no articles meeting our inclusion criterion.

*3.3.4. Data extraction*

We carefully extracted the data to understand the pros and cons of AutoML tools. Part of the process involved collecting pertinent data and evaluating each research article. Here we detail the main data we collected from each article:

**Code**: We introduced a source tracking identification code, which allowed us to easily link the extracted data to the specific article from which it originated. This code served as a critical reference point for the origin of our findings.

**Title**: The title of each article to quickly identify them.

**Year**: We kept track of the year each article was published so we could see how AutoML tools have changed over time and see if any limitations or benefits have become more apparent.

**Extract**: Parts of the articles' text that address the strengths and weaknesses of AutoML tools. Our subsequent qualitative analysis relied on these extracted passages.

**Type of AutoML tool**: We categorized the mentioned AutoML tools as either commercial, open source, a combination of both, or unspecified. This categorization provided valuable context on the accessibility/availability of the tools and their main benefits and challenges.

**Method**: We distinguished between articles that made unsupported claims about AutoML tools and those that provided evidence-based assessments. This differentiation allowed us to assess the rigor and credibility of the information.

**Examples of referenced tools**: We compiled a list of AutoML tools that were referred to or discussed for each article – this list of tools served as a reference point for identifying trends and popular choices in the field.

By capturing this information from each article, we were able to compile a dataset that forms the foundation of our analysis. The extracted data is also available in our online open science repository.

*3.3.5. Data analysis*

To improve our catalog, we conducted a thematic synthesis [43] of the advantages and disadvantages of AutoML tools. More precisely, we assigned codes to each recognized advantage and drawback. Each code was designed to represent snippets with similar characteristics, making it easier to classify them into broader topics. The process of translating codes into themes required careful consideration of how codes could be combined to create comprehensive overarching themes. As we moved further away from the text, the level of abstraction increased, which improved the ability to apply the concepts to a wider range of situations.

This was not a single step; instead, it was an iterative process. During the various iterations, certain codes defined in previous cycles were incorporated into other codes, reclassified, or eliminated. Advancing in the translation process required rearranging and reclassifying encoded material into various, and occasionally innovative, codes. The process ended once the saturation point was reached in identifying the potential themes that emerged from the data. A mind map (Figures 8 and 9) was used to organize the many codes into coherent themes.

The article's first author conducted the initial coding and thematic analysis. The resulting collection of codes and themes was independently peer-reviewed by the other three authors (each one reviewed one-third of the codes) and afterward subjected to collaborative discussion and improvement, including contributions from all authors. The ultimate codes and themes were established by reaching a consensus among the four authors.

*3.4. MLR protocol application*

Our initial query on the Scopus database with the defined search query for the white literature got 274 results. After applying the exclusion criteria, 24 articles remained. Then, we started the first snowballing iteration. For a comprehensive overview of our three snowballing rounds and the systematic application of our exclusion criteria, refer to Figure 5.

After the first round of snowballing, we had 2,620 articles, of which 1,665 were excluded for being duplicated, grey literature (intentionally not collected during the white literature search strategy), or because they did not contain the word "AutoML" in the title, abstract, or keywords sections. We applied the exclusion criteria in this phase for 955 articles and 23 articles remained.

After the second round of snowballing, we had 2,342 articles, of which 1,461 were excluded because they were duplicated, grey literature, or did not contain the word "AutoML" in the title, abstract, or keywords sections. In this phase, we applied the exclusion criteria for 881 articles and 7 articles remained.

After the third round of snowballing, we had 654 articles, of which 540 were excluded because they were duplicated, grey literature, or did not contain the word "AutoML" in the title, abstract, or keywords sections. We applied the exclusion criteria in this phase for 114 articles and as we included no additional studies, snowballing saturation was reached. In total, 54 white literature studies passed our selection phase.

For the grey literature, Google Search retrieved 737,000 results (in May 2023). We analyzed 199 items from this total to check the exclusion criteria, and we stopped searching further after analyzing three consecutive search pages (with ten items per page) with no relevant articles. A total of 41 items passed the selection process. For the grey literature from the Gartner database, we retrieved 184 articles, of which 117 were excluded after checking for duplicates and the exclusion criteria. 67 articles passed this selection process. Therefore, a total of 108 grey literature sources passed our selection phase. Details on the number of search results retrieved from each search engine for the grey literature are reported in Figure 6 and Table 2.

## 4. Results of the Multivocal Literature Review

*4.1. Overview of the selected articles*

In this multivocal literature review, various articles covering different years were included. The increase in the number of publications on AutoML

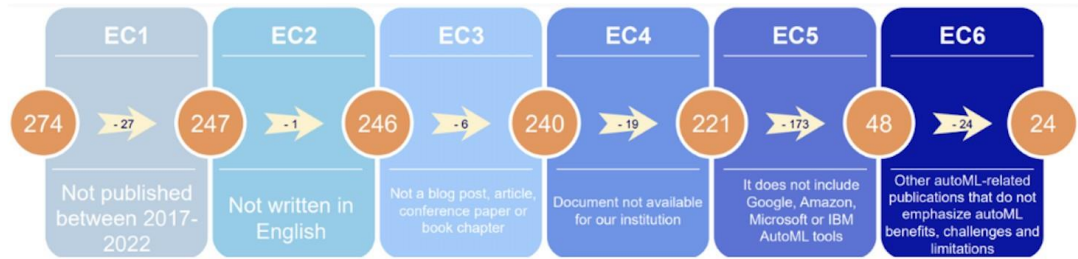
Application of the exclusion criteria in the first Scopus search results.

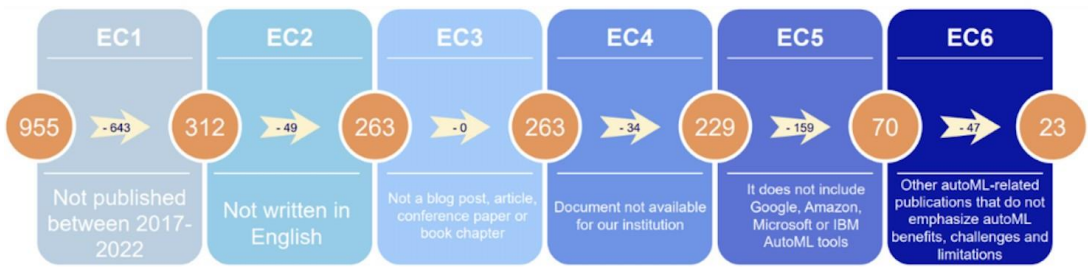
Application of the exclusion criteria in the first snowballing round.

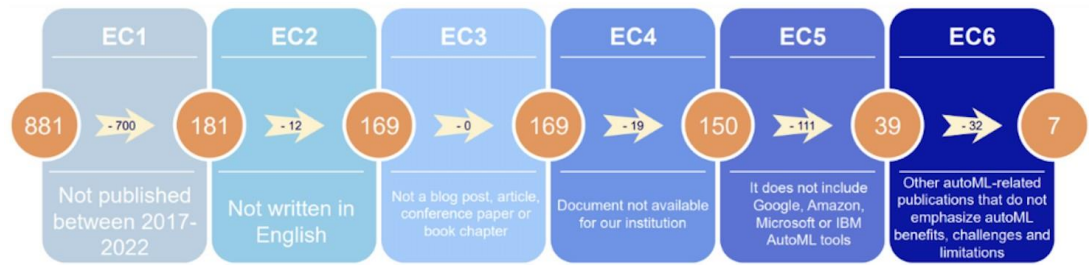
Application of the exclusion criteria in the second snowballing round.

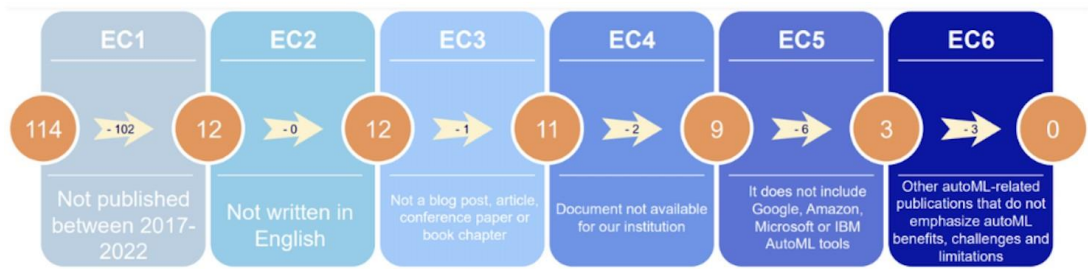
Application of the exclusion criteria in the third snowballing round.

Figure 5: Application of the exclusion criteria in the Scopus search results and snowballing.

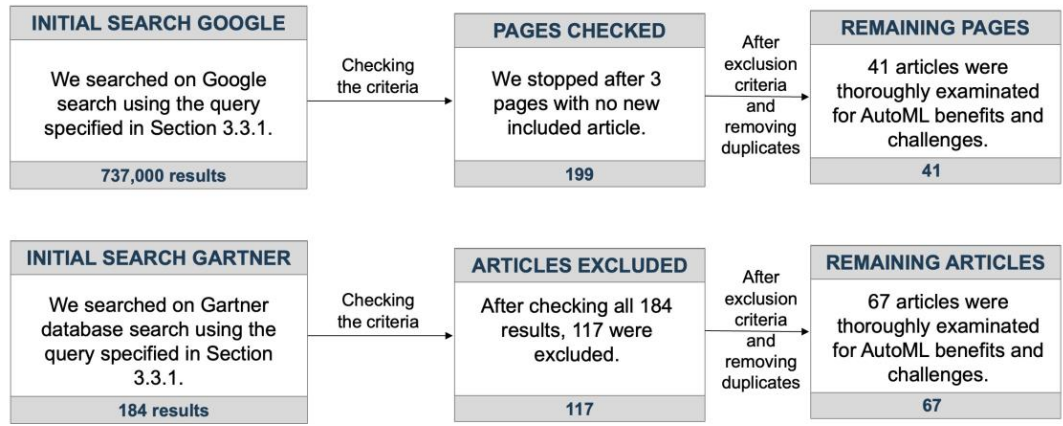

Figure 6: Application of the exclusion criteria in the grey literature (Google Search and Gartner knowledge database).

indicates how popular and useful AutoML tools have become in academia and industry, particularly in recent years. The distribution of the articles analyzed per year is illustrated in Figure 7.

The articles examined cover a wide variety of AutoML application domains, demonstrating how versatile AutoML is across various scenarios. Among the several domains covered, we highlight the following.

**Healthcare**: Many of the reviewed articles are related to healthcare analytics, pathology, and clinical decision-making processes, suggesting a strong emphasis on the use of AutoML for these tasks [44, 24, 17, 45, 46, 25, 21, 47, 48, 49, 50, 20, 47].

**Computer vision**: Demonstrating the application of AutoML algorithms to tasks that involve image analysis, object detection, and picture recognition in different settings [51, 52, 17, 53, 25].

**Manufacturing**: Showing how AutoML can improve production processes, quality assurance, and preventive maintenance [54, 13, 55].

**Water quality**: A domain that showcases the use of AutoML in environmental science, emphasizing data-driven approaches to water quality problems [56].

**Internet of Things (IoT)**: Using AutoML to detect anomalies and per-

**Number of excluded articles for each criterion**

| | Criterion | Excluded Google | Excluded Gartner |
|---|---|---|---|
| **EC1** | Not published between 2017 and 2022 | 89 | 18 |
| **EC2** | Not written in English. | - | - |
| **EC3** | Not a blog post, article, conference paper, or book chapter. | 55 | - |
| **EC4** | Document not available for our institutions. | - | - |
| **EC5** | It does not include Google, Amazon, Microsoft, or IBM AutoML tools. | 9 | 78 |
| **EC6** | Other AutoML-related publications that do not emphasize AutoML benefits, challenges, and limitations. | 5 | 1 |

Table 2: The exclusion criteria and the number of articles excluded from the grey literature.

form predictive analytics in IoT applications, for example, smart grids, intelligent vehicles, smart homes, smart agriculture, and smart healthcare [14].

**Anomaly detection**: Showing how AutoML can be employed for tasks such as fraud detection, intrusion detection, and healthcare system monitoring [57].

**Sentiment analysis**: Demonstrating the versatility of AutoML methodologies to comprehend and evaluate textual data for sentiment-related tasks [58].

*4.2. AutoML benefits*

The codes and themes that emerged from our thematic analysis of the benefits of AutoML are schematically represented in Figure 8. In the following paragraphs, we will go through each theme and provide a description and a few examples for each underlying code.

*4.2.1. Infrastructure*

**Provide scaling efficiency:** AutoML facilitates efficient scaling of machine learning initiatives, making it easy to handle larger datasets and more complex models. As highlighted in [15], Autopilot streamlines the process by offering an automatic hardware recommendation feature. This dynamic allocation of computational resources for each algorithm, dataset, and feature preprocessing pipeline aims to mitigate out-of-memory errors, ensuring seamless operations across various complexities. Moreover, as emphasized by [46], such platforms not only offer scalability, but also allow users to enhance existing models by incorporating more complexities into the classifiers. This scalability is very helpful when changing models to accommodate changing needs and complicated data. Furthermore, as mentioned in [59], AutoML provides users with the flexibility to use the service locally or take advantage of the performance and scalability offered by Azure cloud services. This adaptability allows users to tailor their infrastructure based on specific needs while harnessing the scalability advantages of cloud-based solutions. In the study detailed in [60], Microsoft Azure ML Studio is recognized for its high scalability, deploying instances on the Amazon cloud while also offering on-premise functionality. This multi-faceted scalability ensures seamless operation regardless of the computational environment, facilitating versatility in deployment options for users' diverse needs.

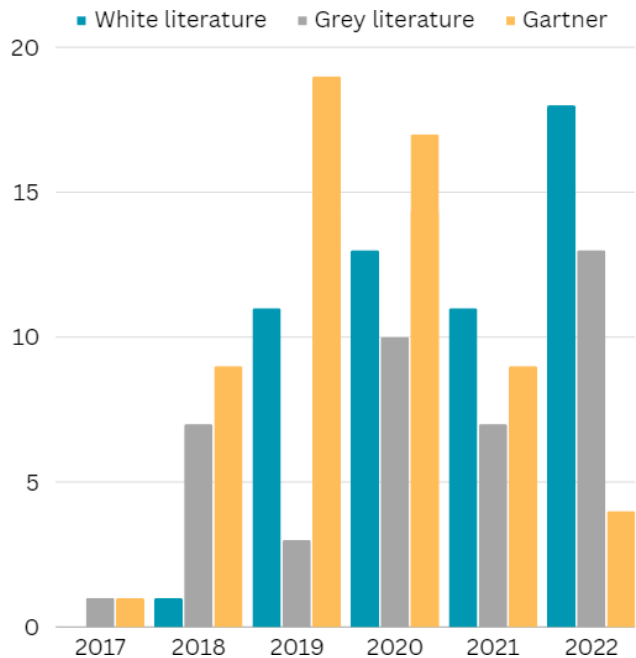

Figure 7: Distribution of the included articles per year.

**Allow easy API integration:** AutoML tools often offer easy integration with various APIs, allowing seamless integration into existing workflows and applications. As detailed in [21], after the model's training phase, its rapid integration into various applications, including online and mobile apps, is achievable without complexity or prolonged setup procedures. This streamlined integration capability enhances the adaptability of AutoML models, enabling their deployment and utilization across a spectrum of platforms and applications.

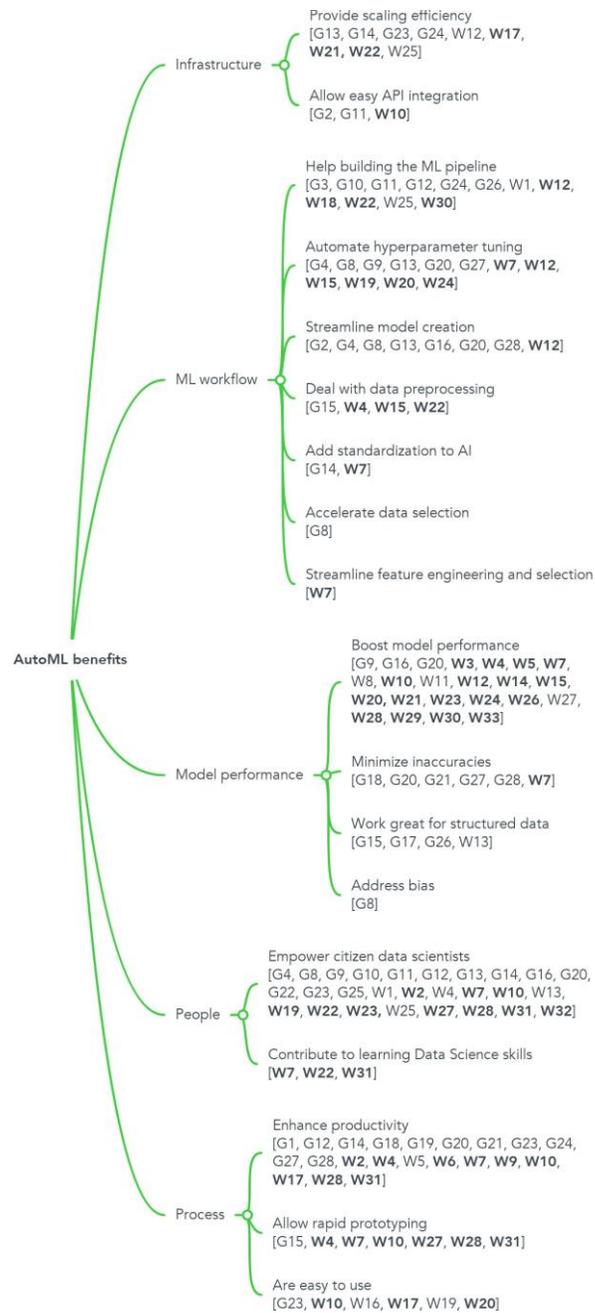

Figure 8: The benefits of using Automated Machine Learning tools. The bold text references are evidence-based extractions. The non-bold references indicate unsupported claims.

*4.2.2. ML workflow*

**Help building the ML pipeline:** AutoML helps build end-to-end machine learning pipelines, from data pre-processing to model deployment. This comprehensive support ensures a systematic and integrated approach to machine learning development. According to [16], AutoML tools not only standardize the ML workflow, but also enhance reproducibility, code maintainability, and knowledge sharing, streamlining collaborative efforts in machine learning projects. For example, platforms such as Amazon SageMaker Studio, as articulated in [15], provide data science teams with a unified web-based visual interface, consolidating all steps of the machine learning process within a single environment. Furthermore, as highlighted in [61], AutoML services play an integral role in the construction of fully integrated MLOps pipelines, reducing the burden on developers to integrate disparate tools and ensuring seamless compatibility among components. This automated pipeline significantly increases developers' productivity, expediting the delivery of new and enhanced application functionalities. As mentioned in [62], AutoML streamlines the complexities associated with building, testing, and deploying novel ML frameworks, thereby simplifying the processes essential for addressing line-of-business challenges. In [21], the use of AutoML demonstrated that ophthalmology residents and fellows lacking coding backgrounds could successfully create deep learning models. Interestingly, the custom models developed by AI specialists performed less effectively than the models generated through AutoML, showcasing the ease of use and efficacy of AutoML for individuals without extensive coding expertise.

**Automate hyperparameter tuning:** AutoML excels in automating hyperparameter tuning, optimizing model configurations for better performance. This is particularly beneficial as selecting appropriate hyperparameters can significantly impact the success of a machine learning model. As cited in [16], participants commonly use AutoML for tasks such as hyperparameter tuning and model selection during the modeling phase. Furthermore, according to [17], AutoML Vision harnesses Google's neural architecture search technology to automatically identify the most effective neural net-work architecture and hyperparameters, streamlining the optimization process. Furthermore, tools such as AutoGluon, highlighted in [45], enable users to define network architectures while enabling thorough parameterization of hyperparameters, offering greater control over model configurations. In a case study assessing the impact of AutoML at GNP Seguros, an insurance company, the AutoML tool achieved an exceptional success rate of 99.2% by adeptly selecting the most suitable classification algorithm and fine-tuning its hyperparameters, as detailed in [18]. Additionally, as articulated in [63], AutoML services automatically improve model performance and accuracy by fine-tuning their hyperparameters, underscoring the significant role that AutoML plays in optimizing machine learning models.

**Streamline model creation:** AutoML streamlines the complex process of developing machine learning models, empowering users to build, validate, and deploy models with minimal manual intervention. As articulated in [63], AutoML services make it easier to create datasets that are used to train machine learning models. When you send these datasets to the services, they instantly compare them with different machine learning methods to find the best one to build an ML model. Furthermore, as highlighted in [64], AutoML tools look at the data and pick the best methods that can be used to build and improve the model.

**Deal with data preprocessing:** AutoML streamlines various data preprocessing tasks, encompassing the management of missing values, feature scaling, and categorical variable encoding. This not only expedites the process, but also guarantees a uniform and error-free input for machine learning models. As highlighted in [13], the use of Auto-sklearn substantially reduced data preparation efforts, requiring only basic data cleaning such as handling NaNs, null columns, and type conversion. Additionally, as per [15], Amazon SageMaker Autopilot adeptly identifies imbalanced binary classification datasets and adjusts the ML pipeline accordingly, resulting in notable improvements in prediction accuracy.

**Add standardization to AI:** AutoML platforms streamline and standardize the machine learning (ML) workflow. According to [16], AutoML *"Standardizes the ML workflow for better reproducibility, code maintainability, knowledge sharing. Another benefit of the black-box nature of Auto-ML tools is that by having a predetermined search space that does not change, there is more standardization of the ML development process, leading to better comparisons across models, code maintainability, and effortless knowledge transfer."*.

**Accelerate data selection:** AutoML accelerates the data selection process by quickly identifying and utilizing relevant datasets. This acceleration is particularly valuable when dealing with large and diverse datasets. According to [65], "*today, AutoML offers the following benefits: it reduces the time to identify the best data sources and hyperparameter tuning settings.*"

**Streamline feature engineering and selection:** AutoML simplifies the feature engineering and selection process, allowing users to identify and incorporate relevant features efficiently. This is crucial for optimizing model performance by focusing on the most influential variables. According to [16], "*feature engineering and feature selection are among the most automated data preprocessing tasks.*"

*4.2.3. Model performance*

**Boost model performance:** AutoML contributes to enhanced model performance by automating the selection of optimal algorithms, features, and configurations, resulting in models that are better suited to the underlying data. As highlighted in [16], many participants, especially ML engineers, emphasize the capability of AutoML tools to rapidly produce superior models, representing a significant advantage. In the study by [20], CFML (Code-free ML) systems outperformed traditional ML object detection systems by using multiple object identification models and selecting the one that exhibits the highest performance. This approach resulted in notable performance gains, underscoring the efficiency of AutoML-driven model selection. Moreover, as detailed in [25], their AutoML Vision model exhibited slight but noteworthy improvements over previously published models, demonstrating the continuous evolution and refinement achievable through AutoML technologies. In another study documented by [21], the AutoML model showcased exceptional performance metrics, boasting values in recall (81%), precision (71%), and F1 score (79%), emphasizing its efficacy in achieving a balance between accuracy and robustness. Furthermore, as mentioned in [18], an accuracy rate of 98.1% was attained using AutoML, surpassing the performance of any manually trained models used previously, underlining the substantial performance enhancements facilitated by AutoML. In a separate investigation by [66], employing IBM Visual Insights with AutoML, precision and recall rates reached 90% and 100%, respectively, showcasing the exceptional accuracy and reliability achievable through AutoML-driven approaches in specific use cases.

**Minimize inaccuracies:** AutoML contributes to reducing inaccuracies by automating processes and minimizing the potential for human error in model development. As stated in [67], AutoML significantly enhances model quality by minimizing the likelihood of inaccuracies due to bias or human errors. By streamlining and automating various stages of model development, AutoML effectively reduces the introduction of biases and errors caused by manual intervention. Moreover, as highlighted in [68], AutoML tools act as a safeguard against errors triggered by human interaction, thus minimizing risks and ensuring greater accuracy throughout the modeling process. By automating repetitive tasks and leveraging standardized procedures, AutoML contributes to the creation of more reliable and precise models, ultimately reducing the potential for inaccuracies arising from human intervention.

**Work great for structured data:** AutoML is well suited for structured data, where its automation capabilities can be leveraged effectively for feature engineering and model building. As noted in [69], AutoML excels particularly in the management of structured data, highlighting its effectiveness in supervised learning tasks, particularly with small to medium-sized datasets characterized by structured formats. The strength of AutoML lies in its ability to rapidly navigate through numerous model alternatives in a remarkably short timeframe. This attribute is highlighted as advantageous for proof-of-concept and prototype development, as mentioned in the same source. By expediting the exploration of diverse model options, AutoML significantly aids in the rapid iteration and validation of concepts, enhancing its utility in early-stage development and demonstrating when dealing with structured data environments.

**Address bias:** Some AutoML tools incorporate features that address bias in machine learning models, contributing to more equitable and fair predictions. As highlighted in [65], these tools streamline the process by recommending the algorithm most suitable for a specific use case while simultaneously addressing algorithmic bias. This multifaceted approach not only optimizes time efficiency, but also actively works towards minimizing biases inherent in machine learning models, contributing to more ethically sound and equitable predictions.

*4.2.4. People*

**Empower citizen data scientists:** AutoML empowers individuals without extensive machine learning expertise, allowing citizen data scientists to harness the capabilities of machine learning for their specific use cases. As highlighted in [26], Google AutoML facilitates the training of a diverse range of machine learning models in various domains, offering a user-friendly experience with minimal technical complexity. According to [19], increased automation enables domain experts to effectively use machine learning technologies. This sentiment is echoed by [20], emphasizing Google AutoML's target users from non-technical backgrounds, enabling professionals to harness the power of ML without necessitating complex scripts or programming. This approach allows them to gain valuable insight while maintaining their domain expertise intact. Moreover, [15] underscores how Autopilot demystifies machine learning for end users lacking expertise in the field, offering a starting point for applying the predictive capabilities of ML to business problems. It helps users understand the tangible value of ML in specific scenarios, bypassing the costlier and riskier alternative of hiring professional data scientists. In a study by [45], classifiers built with AutoML tools performed comparably to the findings in the literature. In particular, these classifiers were created using generic presets with minimal interaction, diverging from prior work that employed task-specific network architectures and optimized training procedures. Additionally, as described in [47], both Google Cloud AutoML and Apple Create ML contribute to the democratization of machine learning. These systems demonstrated robust performance in trained lung and colon cancer diagnostic models, demonstrating no statistically significant differences between them, further illustrating the usability and effectiveness of AutoML platforms for users in various domains.

**Contribute to learning data science skills:** AutoML serves as an educational tool, helping users learn data science concepts and techniques through hands-on experience with automated processes. As highlighted in [16], users who gained access to the search history of AutoML tools reported significant learning experiences. This included insights into new modeling techniques, the implementation of specific ML algorithms, understanding of model architecture, evaluation of model performance across distinct tasks, and understanding of model resource consumption. This first-hand exploration through AutoML not only fosters familiarity with diverse data science techniques, but also provides valuable insights into the practical application of machine learning, contributing significantly to users' skill development in this field.

*4.2.5. Process*

**Enhance productivity:** By automating repetitive and time-consuming tasks, AutoML significantly improves overall efficiency and productivity in the machine learning development process. As noted in [22], people with high technical expertise, such as data scientists, can use AutoML systems to accelerate routine tasks, thus improving the speed and efficiency of their workflows. Furthermore, according to [19], AutoML tools substantially increase the productivity of data scientists by automating a considerable portion of manual work, allowing them to focus on more complex aspects of model development and analysis. In alignment with [56], AutoML plays a vital role in democratizing AI by alleviating redundant human labor, thus improving efficiency in both time and results. This sentiment is echoed in this study [70], which showed that a majority of AutoML solutions consistently delivered reasonable performance across various datasets, underscoring their reliability in achieving efficient outcomes. As emphasized in [16], one of the primary benefits reported by numerous participants, especially ML engineers, is the ability of AutoML tools to accelerate model-building processes while ensuring superior model quality. Moreover, according to [55], AutoML's ability to significantly reduce the time required for data analysis, algorithm training, and optimization further underscores its role in expediting the overall machine learning pipeline, ultimately contributing to improved efficiency and productivity in data-driven tasks.

**Allow rapid prototyping:** AutoML is particularly advantageous in the prototyping phase of an ML project, allowing rapid experimentation and iteration in model development. It can potentially reduce deployment costs by streamlining the development process, making machine learning more accessible without substantial financial investments. According to the findings of [13], Auto-sklearn serves as an initial platform for data scientists, offering both a foundational groundwork for manual implementation and a springboard for further enhancements to their solutions. Additionally, as highlighted in [16], AutoML's influence has significantly reduced the entry barrier by empowering users to develop prototypes fast. These prototypes serve as invaluable tools for assessing the feasibility and potential impact of machine learning applications. In a study conducted by [22], participants actively used AutoML technologies to rapidly prototype and design viable solutions for data preparation and analysis, underscoring its pivotal role in accelerating the ideation phase. Moreover, as mentioned in [55], AutoML plays a dual role by serving as a prototyping tool while seamlessly integrating into various business processes, demonstrating its versatility and applicability across different domains.

**Are easy to use:** With a user-friendly interface, AutoML tools make ma-chine learning accessible to a wider audience, fostering collaboration between domain experts and data scientists. As noted in [20], CFML systems present significant advantages, including ease of use, cost-effectiveness, and a visually intuitive interface that displays true/false positives and negatives. Moreover, these systems enable for easier model deployment for further analysis, promoting accessibility and ease of use. Furthermore, as highlighted in [71], AutoML tools stand out for their user-friendly interfaces, which require minimal machine learning expertise to effectively train models. This characteristic empowers a wider spectrum of users, allowing them to engage in model development and analysis without extensive technical knowledge. Consistently, as mentioned in [17], the user-friendly interface of AutoML Vision shows great potential in clinical practice, helping physicians in decision making processes. The intuitive design and accessibility of such tools contribute significantly to their adoption and usability across diverse domains, ultimately fostering collaboration between domain experts and technical specialists.

*4.3. AutoML limitations*

The codes and themes that emerged from our thematic analysis on the limitations of AutoML are schematically represented in Figure 9. In the following paragraphs, we will go through each theme and provide a description and a few examples for each underlying code.

*4.3.1. Data*

**Have data size constraints:** Challenges arise when dealing with large datasets, as AutoML can face limitations in processing large volumes of data efficiently. As observed in [27], the resolution of large-scale problems remains an ongoing challenge for contemporary AutoML solutions, indicating the persisting hurdles in effectively handling large datasets. Furthermore, as discussed in [14], the application of AutoML models to large-scale datasets proves to be challenging due to the need for multiple training iterations to identify the optimal solution. This underscores the computational complexities inherent in processing vast amounts of data, which often demands extensive computational resources and iterative model training, posing significant challenges for AutoML systems operating at scale. Addressing these constraints requires innovations that enhance AutoML's scalability and efficiency in handling large-scale datasets without compromising performance.

**Do not work well for complex scenarios:** AutoML's effectiveness is constrained in complex scenarios, such as unsupervised learning tasks, where the intricate nature of the data and the absence of labeled examples present challenges. As highlighted in [16], human intervention often compensates for the shortcomings of AutoML, thereby improving its overall performance. However, these limitations become clear when AutoML confronts non-standard use cases and domains, struggling to adapt its predefined frameworks to the unique complexities presented in such scenarios. Furthermore, as indicated in [50], prevailing AutoML approaches exhibit limitations, particularly in managing the scale and diversity of data within biomedical environments. This deficiency underscores AutoML's struggle to cope with the size and diverse nature of datasets, especially within specialized domains like biomedical research, hindering its efficacy in handling the complexities intrinsic to these specific fields.

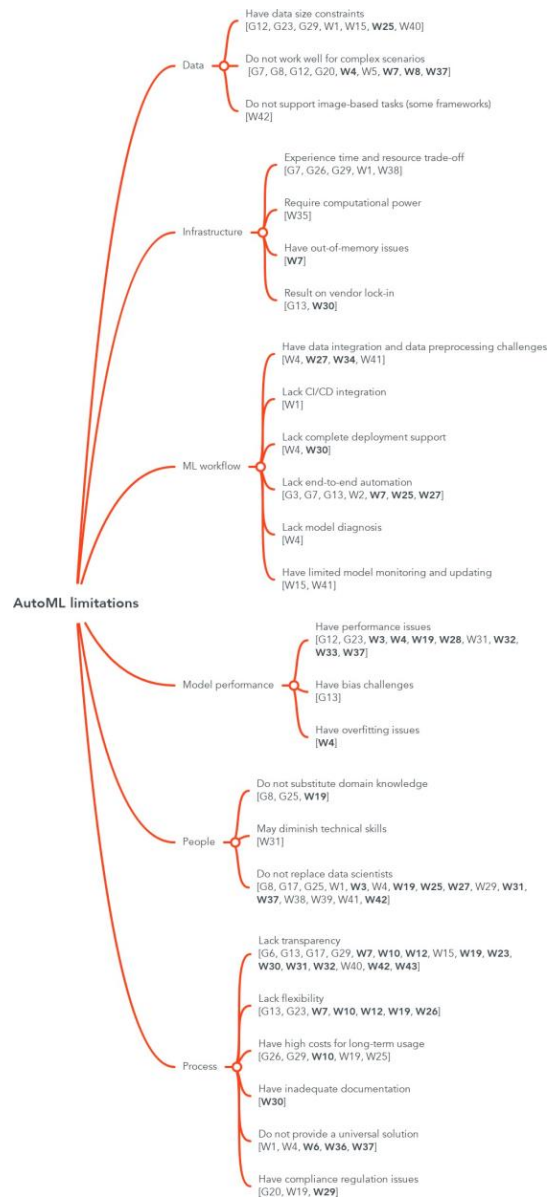

Figure 9: The limitations of using Automated Machine Learning tools. The bold text references are evidence-based extractions. The non-bold references indicate unsupported claims.

**Do not support image-based tasks (some frameworks):** The limited support for image-based tasks poses a challenge, restricting the applicability of AutoML in domains heavily reliant on image data. As highlighted in [72], leading AutoML frameworks still lack comprehensive support for certain image-based tasks, which is especially evident in tasks such as segmentation. Addressing this challenge requires advancements in AutoML frameworks to enhance their capabilities in accommodating and processing diverse image-based tasks, enabling greater applicability across image-centric domains.

*4.3.2. Infrastructure*

**Experience time and resource trade-off:** AutoML processes often involve a trade-off between time and computational resources, necessitating careful consideration of resource allocation to achieve desired results efficiently. As articulated in [26], a larger time budget corresponds to prolonged waiting periods and increased consumption of computing resources, which can result in higher costs, especially when using cloud-based resources. In contrast, a smaller time budget reduces waiting periods, but decreases the likelihood of obtaining the optimal recommendation. This trade-off necessitates strategic decision-making regarding time constraints and resource utilization to strike an optimal balance between achieving efficient outcomes and managing computational expenses.

**Require computational power:** AutoML tools can be computationally intensive, demanding substantial computational power for tasks such as hyperparameter tuning and model training. As emphasized in [73], the execution of AutoML processes requires considerable computational resources to function effectively.

**Have out-of-memory issues:** Resource consumption challenges extend to memory constraints, as large datasets or complex models can lead to memory issues during AutoML processes. As noted in [16], compute-intensive workloads frequently result in system failures, a prevalent concern highlighted by participants utilizing Open-Source Software (OSS) solutions. This issue significantly impacts system performance, contributing to lower ratings in terms of ease of use compared to other tool categories. In this study, participants reported that encountering limitations in main memory capacity constituted one of the primary technical challenges while using AutoML, underscoring the criticality of addressing memory constraints for smoother and more efficient AutoML operations.

**Result on vendor lock-in:** The potential for vendor lock-in emerges as a challenge, with limited interoperability between AutoML solutions, restricting organizations from easily transitioning between platforms and impeding flexibility. As cited in [74], these solutions often introduce common frustrations associated with cloud services, such as a lack of customizability, susceptibility to vendor lock-in, and an opaque operational process.

*4.3.3. ML workflow*

**Have data integration and data preprocessing challenges:** Integrating data from various sources can be a challenge for AutoML, particularly when dealing with heterogeneous datasets that require careful preprocessing and harmonization. As highlighted in [13], the current landscape lacks systems capable of effectively automating the Data Integration phase. This challenge stems from the inherent difficulty of automating tasks involving heterogeneous data sources, such as data synchronization and merging. The complexity arises from the diversity of data sources and the tasks involved in merging different datasets. Successful integration requires a comprehensive knowledge of the types, structures, and complex details of data sources, making automation particularly challenging in scenarios where the nuances of each dataset must be understood and reconciled for effective processing. Thus, addressing the complexities of data integration in AutoML requires advanced capabilities to handle heterogeneous data sources and streamline harmonization processes.

**Lack CI/CD integration:** Integration challenges with continuous integration/continuous deployment (CI/CD) pipelines hinder the incorporation of AutoML into agile development workflows. As observed in [19], AutoML solutions predominantly concentrate on hyperparameter tuning and feature engineering, often neglecting comprehensive considerations from a software engineering and integration standpoint. This lack of holistic integration perspectives from the software engineering angle creates obstacles in aligning AutoML processes within CI/CD pipelines, hindering the efficiency and continuity of agile development workflows.

**Lack complete deployment support:** Deployment challenges arise as AutoML tools may lack support for deploying models in diverse production environments, hindering the transition from development to real-world applications. As highlighted in [75], findings reveal that a significant portion of AutoML practitioners, approximately 13%, encounter difficulties in model load and deployment. This issue is critical as AutoML is expected to streamline model deployment and mitigate stability-related problems. Addressing these challenges requires substantial efforts in future research within Software Engineering to enhance AutoML's capabilities for seamless and reliable model deployment across varied real-world contexts.

**Lack end-to-end automation:** AutoML's focus on specific aspects of the machine learning pipeline results in partial end-to-end support, requiring manual intervention for certain stages of model development and deployment. AutoML tools often prioritize hyperparameter tuning and feature engineering, neglecting other critical aspects of the machine learning pipeline, thus limiting its workflow coverage. As highlighted by [13], existing AutoML systems encompass various ML pipeline steps, but none cover the entire spectrum. Moreover, as mentioned in [16], AutoML focuses mainly on automating model training, leaving users responsible for data pre-processing and post-processing tasks using separate tools.

**Lack model diagnosis:** Diagnosing and understanding model behavior, particularly in complex scenarios, remains a challenge, limiting the interpretability and trustworthiness of AutoML-generated models. As noted in [13], while there are sophisticated approaches for automatic algorithm selection during hyperparameter tuning and training, the comprehensive diagnosis of models remains a challenging area that requires further attention and improvement.

**Have limited model monitoring and updating:** AutoML tools may lack robust mechanisms for model monitoring and updating, crucial for adapting to evolving data patterns and ensuring sustained performance in production environments. As highlighted in [76], while current capabilities focus primarily on the modeling and data analysis stages, there is a notable lack of automation for the labor-intensive and time-consuming phases, such as data preparation or ongoing model runtime monitoring. This gap in AutoML functionalities significantly impacts the efficacy of monitoring model performance in real-time, thereby hindering the automatic adaptation of models to new data trends or shifts in the underlying data distribution. Consequently, the absence of streamlined procedures for continual model updates may impede the responsiveness required to maintain model accuracy and relevance over time. Addressing these limitations in AutoML tools regarding monitoring and updating functionalities is important to ensure the adaptability and sustained efficiency of machine learning models in dynamic real-world scenarios.

*4.3.4. Model performance*

**Have performance issues:** AutoML models may not always achieve optimal performance, as automated processes might lead to suboptimal choices in terms of algorithms, hyperparameters, or feature selection. As evidenced in [13], manual data preparation by data scientists showed superior performance, with a non-tuned Random Forest (RF) outperforming the model generated by the AutoML system, underscoring the limitations of automated processes in certain scenarios. Furthermore, as articulated in [66], specific tasks, such as differentiating between images of viral and bacterial pneumonia, posed considerable challenges for the AutoML model, leading to less favorable outcomes in performance metrics. In another instance, as discussed in [24], automated deep learning models demonstrated limited performance in certain multilabel classification tasks, possibly due to peculiarities within training datasets. These examples underscore the instances where AutoML models may fall short of achieving optimal performance due to challenges in dataset nuances, variations in data structures, and limitations in addressing specific tasks' intricacies through automated processes.

**Have bias challenges:** Despite AutoML's aim to streamline model development, confronting and rectifying biases remains a persisting challenge. Upholding fairness in automated decisions is a crucial objective, but overcoming inherent biases entrenched within training data presents a hurdle. As highlighted in [74], while AutoML providers are taking steps to embed antibias processes and explainability, substantial progress remains essential. The absence of effective measures to address explainability within AutoML platforms poses a significant obstacle. Revealing the model's inner workings and identifying potential sources of suspect outcomes are pivotal in fostering trust and, in case of issues, facilitating remedial actions. The mere involvement of humans does not ensure the absence of bias; therefore, visibility into the model is crucial to mitigate the growing likelihood of biases in decision-making processes. The journey towards mitigating biases within AutoML systems requires further advancements in explainability to fortify transparency, trust, and the ability to rectify biases effectively.

**Have overfitting issues:** Despite automated hyperparameter tuning, AutoML models may still be prone to overfitting, especially when dealing with complex datasets or a limited amount of training data. As noted in [13], the community reports overfitting as an issue associated with Auto-sklearn, often occurring when an excessive amount of time is allocated to the training process. The challenge of overfitting persists despite AutoML's automated optimization efforts, predominantly arising due to complex data structures or limited data samples, emphasizing the need for caution in dedicating excessive training time. Mitigating overfitting issues in AutoML warrants careful consideration of training durations and other strategies to prevent models from excessively tailoring to training data at the expense of generalizability.

*4.3.5. People*

**Do not substitute domain knowledge:** AutoML tools do not substitute the need for domain knowledge, and successful model development still requires a deep understanding of the problem's specific domain and intricacies. As highlighted in [45], the fusion of domain-specific knowledge with machine learning expertise remains indispensable for effectively bridging the gap between a diagnostic task and its assessment metrics and the associated machine learning tasks, encompassing loss and evaluation metrics. Although generic AutoML tools alleviate manual efforts in optimizing Convolutional Neural Network architectures and training procedures, they do not eliminate the essentiality of domain-specific data curation. Detecting and mitigating potential biases and imbalances in datasets continues to be crucial aspects. Moreover, ensuring model generalization beyond the training data requires the use of appropriate data normalization techniques and augmentation strategies. This involves expertise in domain knowledge to ensure the efficacy and applicability of machine learning models in real-world scenarios, going beyond algorithmic optimization.

**May diminish technical skills:** While AutoML streamlines many processes, there is a concern that excessive reliance on automated tools may lead to a decrease in machine learning technical skills. As described in [23], there is skepticism regarding AutoAI's widespread adoption potentially weaken-ing the technical ability of data scientists. Informants stressed their concern about the future generation of data scientists and their proficiency in integrating into an automated data science landscape. This concern revolves around how upcoming professionals might rely heavily on automated data science tools, potentially undermining the development of essential technical competencies in the field. The concern encompasses a potential shift in focus, where reliance on AutoML tools might overshadow the cultivation of in-depth technical knowledge and skills vital for comprehensive understanding and innovation in machine learning. Therefore, while AutoML presents efficiency gains, maintaining a balance that nurtures technical expertise alongside automated tools is crucial for the sustained advancement and evolution of the field of data science.

**Do not replace data scientists:** Contrary to expectations, AutoML tools do not replace the role of data scientists; rather, they complement their work by automating specific tasks. Skilled data scientists remain essential for effective model development and interpretation. As noted in [77], AutoML primarily serves exploratory purposes, while the creation of comprehensive AI systems still largely relies on the expertise of AI professionals. Furthermore, as highlighted in [72], the use of human knowledge significantly reduces the exhaustive computational resources and time required to obtain a high-performance model from the vast search space of AutoML. The practical use of AutoML in real-world settings, as emphasized in [22], requires substantial human effort for effective deployment and utilization. In practice, current AutoML systems, as observed in [76], predominantly target technical personas like data scientists and AI/ML Ops engineers. Despite the automation provided by AutoML, as acknowledged in [78], many real-world scenarios require human supervision. Furthermore, according to [79], designing an effective search space for AutoML requires substantial expertise and knowledge from human experts. A practical search space should combine human prior knowledge with machine learning, considering the trade-off between compactness and encompassing all possibilities. Therefore, while AutoML streamlines certain processes, it is evident that skilled human intervention remains crucial for its effective implementation and optimization in various real-world applications.

*4.3.6. Process*

**Lack transparency:** AutoML's black-box nature poses a significant challenge, as the lack of transparency in its decision-making processes hin-ders the interpretability of models, limiting user understanding and trust in automated outcomes. As expressed in [16], the lack of configurability and transparency is evident in tools such as Google Cloud AutoML. Despite leveraging proprietary Google Research technology to enhance model performance, this platform restricts user input on model types and fails to offer visibility into model internals, impeding users' ability to comprehend model functioning. Similarly, as articulated in [21], the lack of insight into the model architecture and hyperparameters within AutoML tools restricts users from understanding the classification mechanisms or customizing performance parameters, further exacerbating interpretability issues. Furthermore, as highlighted in [45], AutoML Vision lacks logs or reports detailing initialization, network architecture search, or hyperparameter optimization, adding to the opaqueness surrounding model development processes. This opacity, as noted in the same source, signifies a risk in which the internal mechanisms of AutoML tools might undergo imperceptible changes to users, further hindering transparency. Furthermore, as discussed in [75], a substantial percentage of queries from AutoML practitioners revolve around enhancing model performance with limited resources. While AutoML endeavors to abstract model architecture complexities, the need for improved explainability in model predictions remains essential to address practitioners' concerns effectively.

**Lack flexibility:** AutoML's predefined algorithms and workflows may lack the flexibility needed for certain specialized tasks or unconventional problem domains, limiting its applicability in diverse scenarios. As highlighted in [16], customizability ranks notably low among participants' assessments of AutoML tools. Interestingly, the dissatisfaction with customizability stemmed from both insufficient options for customization and an overwhelming degree of flexibility, contributing to the overall lower rating in this aspect. Moreover, as elucidated in [45], AutoML Vision's limitations are evident in its restrictive approach, allowing users to select only one among three presets and specifying a runtime budget based on the dataset size. This lack of expansive customization options restricts users from adequately tailoring the tool to meet the nuanced demands of diverse datasets or specialized tasks. The rigidity within AutoML frameworks poses constraints in accommodating varied needs, leading to dissatisfaction among users due to insufficient options for customization or overly limited choices. These limitations underscore the importance of enhancing flexibility within AutoML tools to widen their applicability across a broader spectrum of problem domains and specific use cases.

**Have high costs for long-term usage:** The adoption of AutoML can incur high costs, both in terms of licensing fees and computational resources required, potentially discouraging smaller organizations with limited budgets. As elucidated in [59], the user-set time budget emerges as a critical parameter in AutoML systems, significantly impacting the system's ability to explore various options within the search space, and thereby increasing the likelihood of receiving superior recommendations. However, a larger time budget amplifies waiting periods and escalates computational resource consumption, directly translating into higher costs, especially when utilizing cloud-based resources. Furthermore, as highlighted in [20], while CFML sys- tems are user-friendly and demand minimal computational expertise, several limitations hinder widespread adoption, particularly among neurosurgeon- scientists. In particular, the cost factor surfaces as a significant concern, wherein continuous usage of the dashboard incurs expenses as billing ap- plies for every hour the model resides in the cloud, regardless of ongoing training or testing activities. These cost considerations underscore a crucial challenge for organizations, especially smaller entities with limited financial resources, as the adoption of AutoML could potentially strain budgets due to the incurred expenses in licensing fees and ongoing computational resource utilization. Addressing these financial implications is imperative for greater accessibility and adoption of AutoML at various organizational scales.

**Have inadequate documentation:** The lack of robust software engineering practices within AutoML processes presents challenges regarding code maintainability, scalability, and collaborative integration, thereby impacting the seamless assimilation of AutoML into established development workflows. As noted in [19], a predominant focus within most existing AutoML solutions revolves around hyperparameter tuning and feature engineering, neglecting comprehensive approaches from a software engineering and integration standpoint. Additionally, as highlighted in [75], similar to empirical studies in the broader ML domain, widespread API misuse arises due to inadequate documentation or insufficient expertise across all phases of the Machine Learning Life Cycle. An illustrative example is an error in API calls for deploying and undeploying Google AutoML's natural language model, revealing documentation inadequacies. This signifies a need for collaboration between researchers in the ML/AutoML field and the Software Engineering domain to address these limitations and enhance documentation practices for improved API utilization. Such collaborative efforts are essential to bridge gaps in software engineering within the AutoML domain, fostering enhanced code management, scalability, and seamless integration into established development practices.

**Do not provide a universal solution:** The absence of a one-size-fits-all solution poses a significant challenge, as AutoML may not be universally applicable across all domains, datasets, and machine learning scenarios. As highlighted in [70], the current landscape lacks a perfect tool capable of consistently outperforming others on a multitude of tasks. This absence of a singularly superior solution underscores the inherent variability in AutoML's effectiveness across different applications. Moreover, as discussed in [49], variability in results persists even within the same dataset, showcasing the absence of a singular method for tasks such as feature extraction, feature selection, algorithm choice, and hyperparameter optimization. This diversity underscores the need to identify the most effective combinations of steps tailored to specific trials rather than relying on a universal approach. Consequently, addressing the challenge of variability in AutoML's effectiveness mandates an agile and adaptive approach to accommodate the unique intricacies of individual datasets and machine learning objectives.

**Have compliance regulation issues:** As noted in [45], cloud-based tools such as AutoML Vision might encounter compatibility issues with local data protection regulations, posing a significant hurdle in ensuring compliance with various regulatory frameworks. Furthermore, as highlighted in [77], a critical concern lies in the documentation of insights obtained during the exploration of AutoML use cases, indicating the need for further attention in this aspect. Moreover, the verification of AutoML applications' adherence to regulatory compliance remains an ongoing challenge, emphasizing the need for more comprehensive measures to ensure alignment with evolving regulatory standards.

In navigating the landscape of AutoML, acknowledging and addressing these challenges are crucial steps toward harnessing its full potential. While AutoML significantly accelerates the machine learning development process, there are some limitations that need to be addressed before it can be effectively used in real-world applications.

## 5. Discussion

*5.1. RQ1 – Advantages of AutoML tools*

Our analysis of the primary advantages of AutoML tools, as addressed in RQ1 ("What are the main benefits of AutoML tools?"), reveals several impactful benefits associated with their use. AutoML tools showcase proficiency in streamlining multiple facets of the machine learning workflow, demonstrating their ability to optimize model performance through various means.

Our study presents a comprehensive list of 18 primary advantages associated with the utilization of AutoML tools. This distinguishes it from the aforementioned systematic literature review (SLR) on AutoML tools by Barbudo *et al.* [36], which explores different research questions, not explicitly focusing on reported benefits.

First, AutoML tools excel at simplifying feature engineering, aiding in the identification and integration of pertinent features critical to enhancing model performance. In addition, these tools expedite data selection, enabling efficient utilization of diverse datasets, which is particularly advantageous when handling extensive and varied data sources.

AutoML significantly contributes to data preprocessing by automating tasks such as handling missing values, scaling features, and encoding categorical variables. This automation not only saves time but also helps to have standardized, error-free data input for models.

Additionally, AutoML simplifies the overall model creation process by automating the building, validation, and deployment stages. It optimizes model configurations by automating hyperparameter tuning, consequently enhancing model accuracy and efficiency.

Beyond technical benefits, AutoML tools are pivotal in democratizing machine learning by offering user-friendly interfaces and automating complex processes, enabling individuals without extensive machine learning expertise to interact with ML models and perform data analysis. Furthermore, these tools incorporate features to mitigate bias in machine learning models, contributing to more reliable predictions.

The cumulative advantages highlighted underscore the substantial impact of AutoML in optimizing the entire machine learning workflow. These tools serve as catalysts for efficiency, accelerating processes, improving model performance, increasing productivity, aiding scalability, and facilitating rapid prototyping, thus broadening the approachability and effectiveness of artificial intelligence methods.

*5.2. RQ2 – Limitations of AutoML tools*

AutoML tools play a crucial role in democratizing machine learning, but their widespread adoption faces several challenges. In response to RQ2 ("What are the main limitations of AutoML tools?"), our analysis of multiple articles uncovered significant obstacles that shed light on the complexities and limitations inherent to AutoML. Unlike this previous SLR on AutoML challenges [35], we focus on AutoML in general, apply a different search strategy for the white literature - including snowballing, complement evidence with grey literature, and follow auditable thematic analysis procedures. As a result, we provide a catalog of 25 limitations of AutoML tools qualitatively organized into themes (Figure 9).

One substantial challenge lies in the lack of transparency within AutoML's decision-making processes, which impacts the understanding and trust in the generated models. In addition, issues such as vendor lock-in, limited interoperability, effective bias mitigation, and a lack of robust software engineering techniques pose important hurdles. The adaptability of established AutoML algorithms for specific tasks is constrained, compounded by high costs and reduced efficacy in complex scenarios. Furthermore, integration difficulties, an excessive focus on specific tasks, and struggles with diverse datasets and unstructured data compound these limitations. Moreover, AutoML's inability to replace domain knowledge, the potential decrease in technical skills due to excessive reliance, and its dependency on high-quality input data further highlight its shortcomings.

These limitations, illustrated in Figure 9, cover various challenges in the ML workflow. These include inadequate integration with Continuous Integration/Continuous Deployment (CI/CD) systems, an overemphasis on hyperparameter tuning and feature engineering, issues with data integration and preprocessing, incomplete deployment support, limited capabilities in model monitoring and updating, and inadequate handling of image-based tasks in certain tools. It is important to note that these characteristics may vary significantly among different AutoML tools, especially if they are open-source or commercial tools.

Moreover, AutoML tools face constraints in replacing human expertise. While these tools enhance performance and efficiency, they cannot completely substitute domain expertise and ML knowledge.

Transparency remains a critical issue, as AutoML tools often lack clarity on their internal mechanisms, leading to difficulties in understanding and trusting their outcomes and decisions.

In addition, challenges persist in managing complex data, handling large datasets, working with unstructured data, evaluating data quality, and integrating diverse data sources. Performance deficiencies, including suboptimal model outputs and potential overfitting, may undermine the reliability and applicability of AutoML-generated models.

In summary, while AutoML simplifies ML pipeline creation, its effectiveness may be limited, especially with certain open source tools that lack extensive features for diverse pipeline creation. It is crucial to note that AutoML does not replace human expertise but enhances efficiency, necessitating skilled users to leverage its capabilities alongside their own expertise and insights.

*5.3. Future directions*

After conducting a multivocal literature review on the limitations of AutoML tools, many important areas for future enhancement and research have been identified, which are detailed in the following lines.

*Scalability and efficient handling of large-scale data.* Addressing the challenges associated with large datasets is crucial. Future advancements should focus on developing innovations that enhance AutoML's scalability and efficiency in processing large-scale data without compromising performance. This includes strategies to optimize computational resources, reduce processing times, and streamline iterative model training on large datasets.

*Enhanced support for complex scenarios and diverse domains.* Enhancements are needed to improve AutoML's adaptability to complex scenarios and unsupervised learning tasks. Tailoring AutoML frameworks to effectively handle the scale and diversity of data in specialized domains, such as biomedical research or other complex fields, is important. Further research should aim to bridge the gap between predefined frameworks and the unique complexities present in diverse domains.

*Expanded capabilities for image-based tasks.* Advancements in AutoML frameworks should prioritize comprehensive support for image-based tasks such as segmentation, enhancing their applicability in domains reliant on image data. Research efforts should focus on augmenting AutoML's capabilities to accommodate and process diverse image-based tasks effectively, especially for open-source tools.

*Optimized resource allocation and infrastructure management.* Future developments should aim to strike a balance between time, computational resources, and costs in AutoML processes. Innovations in resource management strategies, cost-effective use of computational power, and mitigation of out-of-memory issues are essential to streamline AutoML operations.

*Improvements in data integration, preprocessing, and CI/CD integration.* Innovations in automating data integration from heterogeneous sources, robust preprocessing methods, and seamless integration with CI/CD pipelines are crucial. Addressing these challenges requires advanced capabilities to handle diverse data sources and frameworks.

*Comprehensive model deployment support and end-to-end automation.* Enhancements in AutoML should focus on comprehensive support for deploying models across diverse production environments. Moreover, efforts are needed to achieve more significant steps in end-to-end automation, covering all stages of the machine learning pipeline without requiring extensive manual intervention.

*Enhanced model diagnosis, monitoring, and updating.* Future developments should prioritize robust mechanisms to diagnose model behavior, real-time monitoring, and automated updating. Ensuring that models adapt to evolving data patterns is crucial for sustained performance in real-world applications.

*Performance optimization and bias mitigation.* Research efforts should aim to address issues related to suboptimal model performance, overfitting, and effective bias mitigation. The achievement of better model generalization and fairness in decision-making processes remains critical.

*Balancing technical expertise and automation.* Striking a balance between leveraging automation and nurturing technical expertise in data science is important. Future directions should focus on ensuring that AutoML tools complement, rather than replace, the role of skilled data scientists, fostering continuous innovation and understanding in the field.

*Enhanced transparency, flexibility, and cost considerations.* Improving transparency in AutoML decision-making processes, improving flexibility for customization, and addressing cost implications are crucial to broader accessibility and usability across diverse organizations and problem domains.

*Compliance and regulation alignment.* Future AutoML developments should ensure alignment with evolving regulatory standards, facilitating compatibility with various data protection regulations and documentation requirements.

In summary, addressing these specific challenges and prioritizing these future directions would significantly improve the efficiency, applicability, and ethical implementation of AutoML in various real-world scenarios.

## 6. Threats to validity

We aimed to cover a broad spectrum of AutoML content from academic and non-peer-reviewed literature. Therefore, we employed an efficient search strategy for the white literature [38] and described our search strategies for white and grey literature in detail, making the details of each filtering step available in our open science repository [42]. Despite employing comprehensive search strategies and examining multiple sources, there remains the risk of overlooking relevant information. Furthermore, the subjective nature of the exclusion criteria may have influenced the completeness of our data collection. To mitigate this threat, We peer-reviewed the application of the search strategy and the exclusion criteria.

Another threat concerns publication bias. While negative research is important because it demonstrates what does not work, scientific literature has a publication bias toward positive results [80]. Negative research is less cited, published, and commonly considered less scientifically interesting [80]. This threat is inherent to literature reviews and cannot be mitigated. Nevertheless, we still found an expressive number of limitations (25).

Finally, regarding the data analysis, while conducting the thematic synthesis on the advantages and limitations of AutoML tools, we recognize some constraints that could impact the reliability of our findings. In our thematic synthesis, we assigned codes to the benefits and limitations, with the goal of transforming them into overarching themes. This process allowed us to condense a large amount of material into more concise elements. However, this process introduced subjectivity and potential biases. To reduce this threat, we peer-reviewed all the codes and carefully discussed them, involving the entire team of authors. Furthermore, to provide complete transparency and enable auditing of our qualitative coding, we made the coded data available in our open science repository.

## 7. Conclusion

This study investigates the benefits and limitations of AutoML tools identified in the white and grey literature, emphasizing their varied functionalities and effects. The study employed comprehensive and transparent search strategies and qualitative thematic analysis procedures, revealing 18 reported benefits and 25 limitations. We visually organized the benefits and limitations into a catalog grouped by themes using mind maps. In general, AutoML tools streamline machine learning workflows, simplify tasks, empower users, improve model performance, increase efficiency, scalability, and accelerate prototyping. However, there are still ongoing difficulties, including the extent of workflow coverage, the limited ability to replace human expertise, problems with transparency and handling diverse data, and potential performance drawbacks.

While AutoML simplifies the process of creating machine learning pipelines, the effectiveness and breadth coverage of these tools can vary. One key point is that these tools enhance human expertise rather than replace it, requiring skilled users to use their capabilities. Gaining a comprehensive understanding of the AutoML landscape is essential for optimizing machine learning progress and choosing the tools appropriately.

Hence, the findings can be used by practitioners to consider trade-offs between such benefits and limitations to conduct effective evaluations of AutoML solution options. Furthermore, they can be used by researchers to steer future research addressing the current limitations.

# Appendix A. List of papers referenced in the Mind Map

In the following table, we report the list of all the white and grey literature that directly contributed to the mind maps built as part of this study.

**Multivocal literature review - papers included**

|  | **Title** | **Ref.** |
|---|---|---|
| **W1** | Automated Machine Learning: Techniques and Frameworks | [26] |
| **W2** | AutoML to Date and Beyond: Challenges and Opportunities | [19] |
| **W3** | Benchmarking AutoML-Supported Lead Time Prediction | [54] |
| **W4** | Automated machine learning for predictive quality in production | [13] |
| **W5** | Automating water quality analysis using ML and auto ML techniques | [56] |
| **W6** | Towards automated machine learning: Evaluation and comparison of AutoML approaches and tools | [70] |
| **W7** | Whither automl? understanding the role of automation in machine learning workflows | [16] |
| **W8** | Automated machine learning: Review of the state-of-the-art and opportunities for healthcare | [50] |
| **W9** | Application of AutoML in the automated coding of educational discourse data | [81] |
| **W10** | Code-free machine learning for object detection in surgical video: a benchmarking, feasibility, and cost study | [20] |
| **W11** | Automatic Machine Learning: An Exploratory Review | [82] |
| **W12** | Machine Learning-as-a-Service Performance Evaluation on Multi-class Datasets | [60] |
| **W13** | Evaluation of Representation Models for Text Classification with AutoML Tools | [83] |

| | **Multivocal literature review - papers included** | |
|---|---|---|
| | **Title** | **Ref.** |
| **W14** | Setting up an Easy-to-Use Machine Learning Pipeline for Medical Decision Support: A Case Study for COVID-19 Diagnosis Based on Deep Learning with CT Scans | [48] |
| **W15** | IoT data analytics in dynamic environments: From an automated machine learning perspective | [14] |
| **W16** | Automated machine learning for identification of pest aphid species | [71] |
| **W17** | Classification of chest radiographs using general purpose cloud-based automated machine learning: pilot study | [46] |
| **W18** | Deep Active Learning for Computer Vision: Past and Future | [53] |
| **W19** | Evaluating generic AutoML tools for computational pathology | [45] |
| **W20** | Evaluation of the performance of traditional machine learning algorithms, convolutional neural network and AutoML Vision in ultrasound breast lesions classification: A comparative study | [17] |
| **W21** | A machine learning model for detecting invasive ductal carcinoma with Google Cloud AutoML Vision | [25] |
| **W22** | Amazon SageMaker Autopilot: A white box AutoML solution at scale | [15] |
| **W23** | Development of a code-free machine learning model for the classification of cataract surgery phases | [21] |
| **W24** | Automated Machine Learning: The New Wave of Machine Learning | [18] |
| **W25** | Automated machine learning: State-of-the-art and open challenges | [59] |
| **W26** | Google Auto ML versus Apple Create ML for Histopathologic Cancer Diagnosis: Which Algorithms Are Better? | [47] |

| | **Multivocal literature review - papers included** | |
|---|---|---|
| | **Title** | **Ref.** |
| **W27** | Fits and starts: Enterprise use of AutoML and the role of humans in the loop | [22] |
| **W28** | The Potential Of AutoML For Demand Forecasting | [55] |
| **W29** | The impact of AutoML on the AI development process | [77] |
| **W30** | Challenges and Barriers of Using Low Code Software for Machine Learning | [75] |
| **W31** | Human-AI collaboration in data science: Exploring data scientists' perceptions of automated AI | [23] |
| **W32** | Automated deep learning design for medical image classification by health-care professionals with no coding experience: a feasibility study | [24] |
| **W33** | Hands-on with IBM Visual Insights | [66] |
| **W34** | An Empirical Study on the Usage of Automated Machine Learning Tools | [84] |
| **W35** | MLOps - Definitions, Tools and Challenges | [73] |
| **W36** | Automated machine learning for healthcare and clinical notes analysis | [49] |
| **W37** | Evaluation of Sentiment Analysis based on AutoML and Traditional Approaches | [58] |
| **W38** | Techniques for Automated Machine Learning | [79] |
| **W39** | A Human-in-the-loop Perspective on AutoML: Milestones and the Road Ahead | [78] |
| **W40** | Automated Machine Learning - a Brief Review at the End of the Early Years | [27] |
| **W41** | How Much Automation Does a Data Scientist Want? | [76] |
| **W42** | Human behavior in image-based Road Health Inspection Systems despite the emerging AutoML | [72] |

| | **Multivocal literature review - papers included** | |
|---|---|---|
| | **Title** | **Ref.** |
| **W43** | Trust in AutoML: Exploring Information Needs for Establishing Trust in Automated Machine Learning Systems | [85] |
| **G1** | Top 10 Data and Analytics Technology Trends That Will Change Your Business | [86] |
| **G2** | Predicts 2020: Artificial Intelligence — the Road to Production | [63] |
| **G3** | Predicts 2019: The Democratization of AI | [87] |
| **G4** | Market Guide for Hosted AI Services | [64] |
| **G5** | Best Automatic Machine Learning (AutoML) Frameworks In 2022 | [88] |
| **G6** | Can automated machine learning outperform handcrafted models? | [89] |
| **G7** | Introduction to Automated Machine Learning (AutoML) | [90] |
| **G8** | Hype Cycle for Open-Source Software, 2020 | [65] |
| **G9** | Hype Cycle for Data Science and Machine Learning, 2020 | [91] |
| **G10** | Magic Quadrant for Cloud AI Developer Services | [41] |
| **G11** | Critical Capabilities for Cloud AI Developer Services | [92] |
| **G12** | Automated Machine Learning — Azure automl — Pros And Cons | [93] |
| **G13** | AutoML: the Promise vs. Reality According to Practitioners | [74] |
| **G14** | Major Benefits of Using AutoML over Hand Coding | [94] |
| **G15** | AutoML: How to Automate Machine Learning With Google Vertex AI, Amazon SageMaker, H20.ai, and Other Providers | [95] |
| **G16** | Automated Machine Learning (AutoML) Libraries for Python | [96] |

**Multivocal literature review - papers included**

|     | Title | Ref. |
| --- | --- | --- |
| **G17** | Text Classification using AutoML Tables — Google Cloud Platform | [97] |
| **G18** | AutoML - The Future of Machine Learning | [98] |
| **G19** | IBM Wants To Make Artificial Intelligence Fair And Transparent With AI OpenScale | [99] |
| **G20** | What Is Automated Machine Learning? Going Beyond The Primer | [100] |
| **G21** | Why Does Your Business Need AutoML? | [101] |
| **G22** | How AutoML is helping to bridge the AI skills gap | [102] |
| **G23** | SAP Sales Forecast using Automated Machine Learning Azure AutoML No-code/Low-code on Microsoft Power BI | [103] |
| **G24** | Automated Machine Learning: What's Its Role in the Future of Work? | [104] |
| **G25** | Auto-Keras and AutoML: A Getting Started Guide | [105] |
| **G26** | What is AutoML and how can it be applied to practice? | [106] |
| **G27** | Pros and Cons of Automated Machine Learning | [107] |
| **G28** | How Automated Machine Learning Addresses the Challenges of Traditional ML Models? | [108] |
| **G29** | Automated Machine Learning (AutoML) with BigQuery ML. Start Machine Learning easily and validate if ML is worth investing in or not | [109] |
| **G30** | Hype Cycle for Supply Chain Planning Technologies, 2020 | [110] |
| **G31** | Hype Cycle for Supply Chain Planning Technologies, 2019 | [111] |
| **G32** | Hype Cycle for Open-Source Software, 2021 | [65] |

Table A.3: List of the articles included in the Multivocal literature review and their references.